\documentclass[prb,aps,twocolumn,showpacs]{revtex4}
\usepackage{color,amsfonts,amsmath,bm}
\usepackage[OT4]{fontenc}

\usepackage{graphicx}
\usepackage{epstopdf}

\begin{document}

\title{Tunneling transfer protocol in a quantum dot chain immune to inhomogeneity} 

\author{Kamil Korzekwa}
 \email{Korzekwa.Kamil@gmail.com} 
\author{Pawe{\l} Machnikowski}
 
\affiliation{Institute of Physics, Wroc{\l}aw University of
Technology, 50-370 Wroc{\l}aw, Poland}

\begin{abstract}

We propose a quantum dot (QD) implementation of a quantum state transfer channel. The proposed channel consists of N vertically stacked QDs with the nearest neighbor tunnel coupling, placed in an axial electric field. We show that the system supports high-fidelity transfer of the state of a terminal dot both by free evolution and by adiabatic transfer. The protocol is to a large extent insensitive to inhomogeneity of the energy parameters of the dots and requires only a global electric field.
\end{abstract}

\pacs{73.21.La, 03.67.Hk}

\maketitle

\section{Introduction} 
\label{sec:intro}

Quantum state transfer in chains of coupled few-level systems (usually referred to as \textit{spin chains})  \cite{bose07a} attracts much attention because of its importance for short-distance quantum information transfer in future quantum information processing devices. One expects that such transfer should take place with minimal external control in a chain that does not require precise engineering of the properties of the individual systems and couplings between them. Moreover, it is desirable to be able to perform transfer on demand at a specific instant of time.  While many formally elegant protocols for quantum state transfer have been proposed \cite{bose07a}, based, e.g., on specific engineering of the couplings \cite{christandl04} or encoding of the qubit in several spins \cite{osborne04}, few of them satisfy these requirements of minimal control and flexibility of physical properties.

Among real systems that can be used to implement the formal idea of quantum state transfer, chains of quantum dots (QDs) seem to be one of the most feasible options \cite{nikolopoulos04}. Electronic coupling between closely stacked dots has been demonstrated \cite{bayer01} and long chains of QDs can be formed using self-assembled growth \cite{oshima08}. The difficulty here is, however, that QDs show considerable uncontrollable inhomogeneity of the confinement energies. Moreover, the tunnel coupling between the neighboring dots cannot be engineered with a high precision.

In this paper, we propose a protocol of quantum state transfer using a chain of stacked QDs which does not depend on perfect homogeneity of energy levels and inter-dot coupling. The only assumption is that the confined states in the terminal dots are energetically shifted with respect to others, which only requires control of the QD properties with an experimentally feasible precision on the order of tens of milli-electron-Volts. The second element is a controllable global electric field that has to be applied along the chain. We show that in such a system the charge state can be transferred between the terminal dots both as a result of free evolution as well as by an adiabatic transition driven by electric field sweep.

The paper is organized as follows. In Sec.~\ref{sec:model}, we describe the system and our methods of simulation. Next, in Sec.~\ref{sec:transfer}, we discuss the state transfer in the QD chain. Finally, Sec.~\ref{sec:concl} concludes the paper.

\section{Model and method} 
\label{sec:model}
\
We consider a chain of $N$ vertically stacked QDs with the nearest neighbour tunnel couplings $J_l$, placed in an axial electric field $\mathbf{\mathcal{E}}$. We assume that the chain is engineered in such a way that the electron energies in the terminal QDs, $E_1^{(0)}$ and $E_N^{(0)}$, are higher than those in the QDs inside the chain. We investigate the situation when the QD chain is doped with one electron. The Hamiltonian of the system is then given by:

\[H=\sum_{l=1}^{N} (E_l^{(0)}+\mathcal{E}l\Delta x )|l\rangle\!\langle l|+\sum_{l=1}^{N-1} J_l\left(|l\rangle\!\langle l+1| +\mbox{h.c.}\right),\]
where $|l\rangle$ denotes the state with the electron in $l$th dot. We include the inhomogeneity of the QD chain by using normal distribution for the values of tunnel couplings and electron energy levels,
\begin{eqnarray}
J_l&\sim&\mathcal{N}(J,\sigma_J^2), \nonumber\\
E_1^{(0)},E_N^{(0)}&\sim&\mathcal{N}(E_T,\sigma_E^2),\nonumber\\
E_l^{(0)}&\sim&\mathcal{N}(0,\sigma_E^2),~ l=2,\dots, N-1.\nonumber
\end{eqnarray}

The evolution of the occupations of the QDs was obtained in two different ways. In the case of free transfer (for time-independent electric field) numerical diagonalization of the Hamiltonian was done in order to find the eigenvalues and eigenstates. Then the initial state $|1\rangle$ (electron on the initial QD) was decomposed into eigenstates of the Hamiltonian $|\Psi_l\rangle$ and evolved in the standard way: $|\Psi_l(t)\rangle=|\Psi_l(0)\rangle\exp(-iE_lt/\hbar)$ to get the final state. In the adiabatic transfer case (when the external field depends on time), numerical solution of the Schr\"odinger equation using the Runge-Kutta method was used.

In our simulations, we use two sets of parameters: for homogenous and inhomogenous QD chains. In both cases $E_T=50$ meV, $J=10$ meV and for the inhomogenous chain we set $\sigma_E^2=10$ meV, $\sigma_J^2=1$ meV (in the homogenous case $\sigma_E^2=\sigma_J^2=0$).

\section{Quantum state transfer} 
\label{sec:transfer}

\subsection{Free transfer}
\label{sec:free}

\begin{figure}[tb]
\begin{center}
\includegraphics[width=8.5cm]{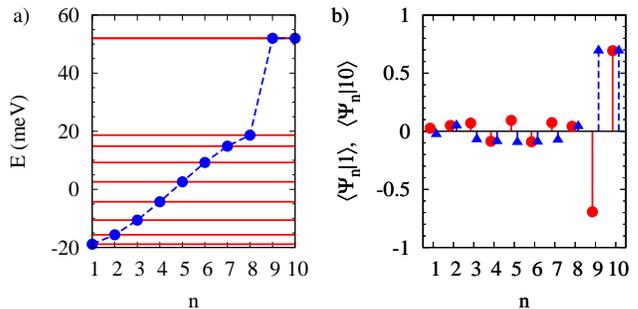}
\end{center}
\caption{\label{fig:Spectrum} Single-electron states in a homogenous QD chain of length $N=10$. a) Energy spectrum; b) Projection of the states with the electron on the initial (solid line, circles) and final (dashed line, triangles) QD on the energy eigenstates.} 
\end{figure}

\begin{figure}[tb]
\begin{center}
\includegraphics[width=8.5cm]{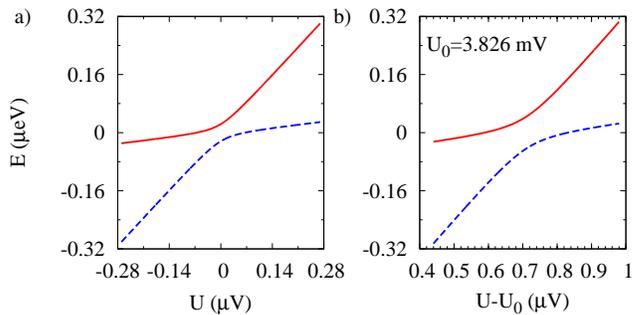}
\end{center}
\caption{\label{fig:resonance} Dependence of energy levels of  $|\Psi_N\rangle$ and $|\Psi_{N-1}\rangle$ on the voltage $U$ between initial and final QDs. a) Homogenous QD chain of length $N=10$; b) Inhomogenous QD chain of length $N=10$.}
\end{figure}

\begin{figure}[tb]
\begin{center}
\includegraphics[width=8.5cm]{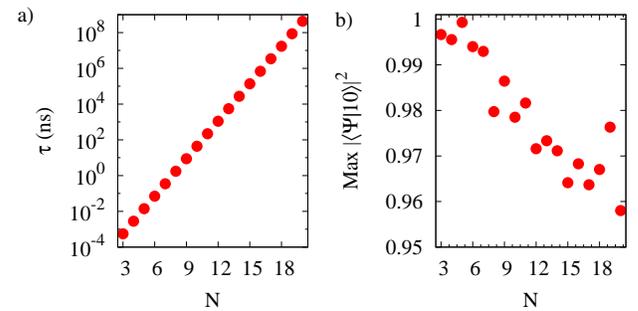}
\end{center}
\caption{\label{fig:tunnel_time} Free evolution transfer time (a) and maximum achieved fidelity (b) as a function of the chain length for a homogeneous chain.}
\end{figure}

\begin{figure}[tb]
\begin{center}
\includegraphics[width=8.5cm]{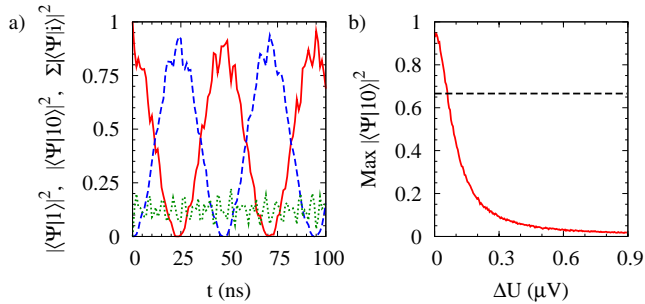}
\end{center}
\caption{\label{fig:free_transfer} Simulation results for a single realization of the free transfer. a) Evolution of the occupation of the initial QD (red solid line), the final QD (blue dashed line) and the QDs inside the chain (green dotted line) for an inhomogenous QD chain with $N=10$ under exact compensation with electric field; b) Maximum obtained fidelity (red solid line) as a function of the deviation from the exact resonance voltage for an inhomogenous chain of length $N=10$. Black dashed line shows the classical limit of efficiency of quantum transfer $F=2/3$ (Ref.~1).}
\end{figure}

Because of higher electron energies on the initial and final dots, the two eigenstates $|\Psi_N\rangle$ and $|\Psi_{N-1}\rangle$ are energetically separated from the others and the corresponding states are localized on terminal QDs (Fig.~\ref{fig:Spectrum}). The states with the electron on terminal QDs are indirectly coupled via the QD chain. The inhomogeneity of energies and couplings modifies the structure of the states delocalized in the central part of the chain, which affects the effective coupling between the terminal states. As a result it shifts the resonance between $|\Psi_N\rangle$ and $|\Psi_{N-1}\rangle$ (the point at which $E_N\approx E_{N-1}$), but only weakly changes its width (Fig.~\ref{fig:resonance}). The two states can be brought to resonance by applying an external electric field which compensates the shift of the resonance. At resonance the free evolution of the system (restricted, to a good approximation, to the subspace $\{|\Psi_N\rangle,|\Psi_{N-1}\rangle\}$) corresponds to electron oscillation between the terminal QDs. The width of the resonance $2V$, which determines the oscillation period, critically depends on $N$, which is reflected in transfer time $\tau_f=\pi\hbar/(2V)$ [Fig.~\ref{fig:tunnel_time}(a)]. Since the width of the resonance depends weakly on the inhomogeneity (as long as it is not too strong) the free transfer time for compensated inhomogenous chain (i.e. with the states $\Psi_N$ and $\Psi_{N-1}$ brought to the resonance with the applied electric field) is of the order of the free transfer time in the homogenous chain. The fidelity depends mainly on the degree of localization of $|\Psi_N\rangle$ and $|\Psi_{N-1}\rangle$ on the terminal QDs, so we can introduce the transfer fidelity parameter $\Delta=1-|\langle\Psi_N|1\rangle|^2-|\langle \Psi_{N-1}|1\rangle|^2$. In our model it is independent of the chain length, hence the transfer fidelity is expected to weakly depend on the chain length, which is confirmed by simulations [Fig.~\ref{fig:tunnel_time}(b)]. The exemplary evolution of the occupation of the QDs in the compensated inhomogenous chain is shown in Fig. \ref{fig:free_transfer}(a). We found out, however, that the transfer fidelity $F=|\langle\Psi_{\mathrm{fin}}|N\rangle|^2$ (where $|\Psi_{\mathrm{fin}}\rangle$ denotes the final state of the system) is very sensitive to small variations of the compensating field, i.e. a small deviation of electric field results in big decrease of the maximum fidelity obtained [Fig.~\ref{fig:free_transfer}(b)].
\subsection{Adiabatic transfer}
\label{sec:adiabatic}
Obtaining free transfer seems to be rather demanding because of the requirement of a very precise control of the electric field. To overcome this problem we propose an adiabatic protocol. The idea is to slowly change the electric field in order to sweep the energy levels of $|\Psi_N\rangle$ and $|\Psi_{N-1}\rangle$ through the resonance. We use an effective 2-level model including the states $|\Psi_N\rangle$ and $|\Psi_{N-1}\rangle$ with the coupling $V$ given by the half of the energy splitting at the resonance. We use the Landau-Zener formula \cite{wittig05} for nonadiabatic tansition probabilities:

\[P_{\mathrm{na}}=\exp\left(-\frac{2\pi}{\hbar}\frac{|V|^2}{\alpha}\right),\]
which yields the dependence of the speed of electric field sweep $\alpha=dE/dt$ as a function of the desired fidelity $F=1-P_{\mathrm{na}}$. 

\begin{figure}[tb]
\begin{center}
\includegraphics[width=8.5cm]{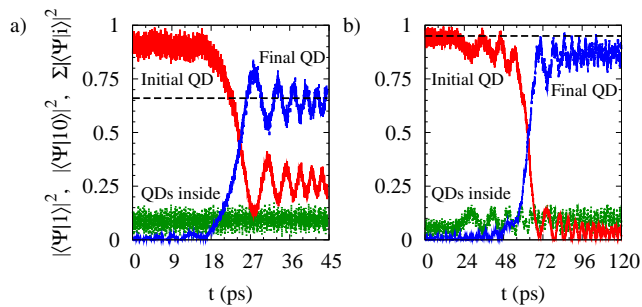}
\end{center}
\caption{\label{fig:adiabatic_transfer} Evolution of the occupation of the initial QD, the final QD and the QDs inside the chain for an inhomogenous chain of length $N=5$. Horizontal dashed lines show the fidelity obtained from the Landau-Zener formula. a) $F=0.66$, $\beta=20$; b) $F=0.95$, $\beta=20$.} 
\end{figure}

\begin{figure}[tb]
\begin{center}
\includegraphics[width=8.5cm]{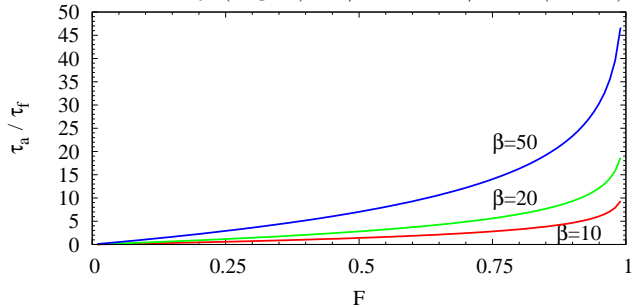}
\end{center}
\caption{\label{fig:ratio} Adiabatic transfer time to free transfer time ratio
for an inhomogenous chain of length $N=5$.} 
\end{figure}

In order to achieve a finite transfer time we narrow  the limits of the electric field sweep to the area where the energy separation of the states is smaller than $\beta V$ for a certain parameter $\beta$ (we assume that for $\Delta E>\beta V$ interaction is negligible). The adiabatic transfer time obtained in this way is $\tau_{\mathrm{a}}=-[\hbar\beta/(\pi V)]\ln\left(1-F\right)$. The Landau-Zener result for the effective two-level model shows good agreement with the simulation of the evolution of the full system (where we sweep $-\beta V<\mathcal{E}<\beta V$ with constant speed $\alpha$) for $\beta$ large enough (see Fig.~\ref{fig:adiabatic_transfer}). In simulations we have not observed significant improvement in the obtained fidelity for $\beta>20$. The ratio of adiabatic transfer time to free transfer time for given $\beta$ depends only on the desired fidelity (Fig.~\ref{fig:ratio}), $\tau_{\mathrm{a}}/\tau_{{\mathrm{f}}}=-2\beta/\pi^2\ln\left(1-F\right)$.

\section{Conclusions}
\label{sec:concl}

We have shown that quantum state transfer is possible in an inhomogeneous QD chain with spectrally shifted terminal dots with a global electric field as the only external control. The proposed protocol is resilient to the QD chain inhomogeneity and the fidelity of the transfer weakly depends on the chain length. Moreover, it is possible to sweep the energies of the two terminal states through resonance using a variable electric field in order to induce an adiabatic passage of the electron between the chain ends, which makes it possible to achieve state transfer on demand. This compares favorably with the earlier proposal \cite{wojcik05} where decoupling of the terminal states 
was due to weak coupling between the terminal QDs and the remaining part of the chain. In that case, it was essential that the terminal states are located in the narrow gap in the spectrum of a finite homogeneous chain which makes the resulting state transfer sensitive to perturbations of the perfect homogeneity and excludes the possibility of sweeping the field over a sufficiently wide range to achieve transfer on demand.

\textbf{Acknowledgment:} This work was supported by the Foundation for Polish Science under the TEAM programme, co-financed by the European Regional Development Fund.

\bibliographystyle{prsty}

\end{document}